# AI, orthogonality and the Müller-Cannon instrumental vs general intelligence distinction

Olle Häggström[1]

Draft, September 14, 2021

**Abstract**

The by now standard argument put forth by Yudkowsky, Bostrom and others for why the possibility of a carelessly handled AI breakthrough poses an existential threat to humanity is shown through careful conceptual analysis to be very much alive and kicking, despite the suggestion in a recent paper by Müller and Cannon that the argument contains a flaw.

**1. Introduction**

Does the possibility of a future AI breakthrough gone wrong pose an existential threat to humanity? In a recent contribution to this topic, Müller and Cannon (2021) point out an ambiguity in the meaning of intelligence which they think constitutes a flaw in the standard argument for existential risk from AI, which they formalize as follows.

> Premise 1: Superintelligent AI is a realistic prospect, and it would be out of human control.
>
> Premise 2: Any level of intelligence can go with any goals.
>
> Conclusion: Superintelligent AI poses an existential risk for humanity.

Let's call this **the AI xrisk argument**. It serves as a kind of cornerstone for the motivation behind much of the growing research field of **AI Alignment**, whose aim is to figure out ways to make sure that advanced AI is equipped with goals that are well-aligned with our desire for human flourishing; se, e.g., Everitt et al (2018) and Hubinger (2020) for technical overviews, or Christian (2020) for a popular account.

Premise 1 of the AI xrisk argument goes back at least to Turing (1951), who famously wrote that "it seems probable that once the machine thinking method had started, it would not take long to outstrip our feeble powers" and that "at some stage therefore we should have to expect the machines to take control". Premise 2 was first formalized by Bostrom (2012) as the so-called **orthogonality thesis**, but appears more implicitly in Yudkowsky (2008) along with the argument as a whole. Bostrom (2014) can be seen as a book-length elaboration on the argument, which has since appeared more or less explicitly in many other contributions to the AI risk literature, including some of my own (Häggström 2016, 2018, 2021), and it seems safe to say that the AI xrisk argument as formalized by Müller and Cannon captures an important aspect of the way many researchers in the field think about existential risk to humanity caused by the wrong kind of AI breakthrough.

But the devil is in the details. Müller and Cannon note that for the argument to be sound, Premise 1 and Premise 2 need to be about the same notion of intelligence, and they find that this is not the case: Premise 1 is about **general intelligence**, and Premise 2 is about what they call **instrumental intelligence**. General intelligence – a term that most often comes up in abbreviated form in the notion of AGI – is used to denote a collection of cognitive abilities that matches or exceeds human level in all relevant aspects, while superintelligence is more of the same: Müller and Cannon quote Bostrom (2014) who says that "superintelligence [is] any intellect that greatly exceeds the cognitive performance of humans in virtually all domains of interest".

---
[1] Department of Mathematical Sciences, Chalmers University of Technology, Gothenburg, Sweden, olleh@chalmers.se

Instrumental intelligence, on the other hand, is about the ability to reason in the service of whatever goal one has. Here Müller and Cannon quote Legg and Hutter (2007) saying that "intelligence measures an agent's ability to achieve goals in a wide range of environments", and Russell (2019) saying that "roughly speaking, an entity is intelligent to the extent that what it does is likely to achieve what it wants, given what it has perceived".

After their diagnosis about different intelligence concepts in Premises 1 and 2, Müller and Cannon go on to see whether the AI xrisk argument can be repaired, either by changing Premise 1 to be about instrumental intelligence, or by changing Premise 2 to be about general intelligence. Their tentative verdict is that it can't be done, and that the AI xrisk argument is therefore simply wrong.

This kind of conceptual analysis can bring valuable clarity to an important discussion. The purpose of the present paper is to build further on the work of Müller and Cannon, and to argue that the only kind of intelligence that is needed in the AI xrisk argument, including both of its premises, is instrumental intelligence. Some issues regarding the argument will remain unsettled in connection with the possibility of moral realism, but this will not be enough to undermine my final conclusion that the AI xrisk argument is still very much worth taking seriously.

## 2. Intelligence and the capability of ethical reasoning

In an oft-cited prototypical example of the kind of scenario that the AI xrisk argument warns might take place, an AI programmed with the goal of maximizing the production of paperclips achieves superintelligence, and it promptly goes on to exploit its supreme intelligence to turn the entire world, including every single human, into a monstrous heap of paperclips. But what kind of intelligence are we talking about here – general or instrumental? If it is instrumental intelligence, then a sufficiently high level of such intelligence is just what the AI needs to achieve its goal of maximizing paperclip production, because achieving goals is exactly what instrumental intelligence is defined to be good at.

However, as Müller and Cannon (2021) point out, the definition of superintelligence is a strengthening of the general intelligence concept, as made explicit in the Bostrom quote in Section 1 above, stating that superintelligence involves "greatly [exceeding] the cognitive performance of humans in virtually all domains of interest". Since ethical reasoning is clearly a domain of interest, the superintelligent AI must be capable of ethical reasoning at a level greatly exceeding that of humans. Surely (or so goes the argument by Müller and Cannon) such an ethically capable agent would figure out how unethical it would be to kill all humans for the sake of producing paperclips, and would change its goals to something more benign than the single-minded focus on maximizing the production of paperclips.

So here, in a nutshell, is the problem Müller and Cannon see with the AI xrisk argument: For anything like the paperclip apocalypse to happen, the AI needs to have only instrumental intelligence at a sufficiently high level, but since it is superintelligent, it will also be able to reason about goals and ethics, which is enough to avert catastrophe.

And yet, I and others in the xrisk and AI safety communities insist on the AI xrisk argument, and that something along the lines of the paperclip apocalypse might very well happen. So what is going on here? In an attempt to understand how we think about this, Müller and Cannon raise four distinct possibilities:

> (I) One might argue that intelligent agents […] are actually *unable* to reflect on goals. (II) Or that intelligent agents *are able* to reflect on goals, but *would* not do so. (III) Or that they would never *revise* goals upon reflection. (IV) Or that they would reflect on and revise goals but still not *act* on them. [Roman numerals added by me.]

Let me go through them one by one. Case (I) would contradict the definition of superintelligence, so we can immediately rule that one out. Case (IV) contradicts the very meaning we attach to the idea of AI goals, which is what the AI acts towards achieving; going all the way back to Turing, the theory of AI is very much an operational and functionalistic one. So case (IV) can be dropped as well. This leaves the more interesting cases (II) and (III) to consider.

As to case (II), let me first note that there is one sense in which a superintelligent agent will be very likely to reflect on goals. To see this, consider the distinction between final and instrumental goals: a final goal is one that an agent strives towards for its own sake, whereas an instrumental goal is one that is pursued merely as a stepping stone for reaching the final goal. To figure out which instrumental goals to pursue in order to efficiently achieve a given final goal is a hallmark of intelligence, and so a superintelligent AI is likely to engage in reflection on such goals. According to the so-called instrumental convergence thesis, there are a number of instrumental goals that a sufficiently intelligent AI is likely to set up pretty much regardless of what its final goal is, some examples of such convergent instrumental goals being self-preservation, self-improvement and resource acquisition (Omohundro, 2008; Bostrom, 2012, 2014). Together with the orthogonality thesis, the instrumental convergence thesis support a view of superintelligent agency that I've dubbed the Omohundro-Bostrom theory of AI goals (Häggström, 2016, 2019). Besides the convergent instrumental goals, there is also an even richer class of instrumental goals whose efficiency towards the final goal are more highly dependent on what this final goal is, as well as on the exact environment and circumstances the agent finds itself in – giving all the more reason for the agent to reflect on them.

But these instrumental goals are not what Müller and Cannon are talking about: the primary thing of interest here is reflection on final goals. And while, as we shall see in the discussion of case (III) below, there are special cases where a superintelligent AI would have very good reason to engage in such reflection, it seems highly uncertain to what extent it will do so under more ordinary circumstances. Still, if we allow ourselves a bit of anthropomorphic reasoning, human experience seems to suggest that it is useful to sometimes zoom out and consider one's life, ambitions and so on from a more detached outside perspective. We certainly seem more prone to do so than cognitively lesser animals, and extrapolating wildly to higher intelligence levels might suggest that a superintelligent AI would do even more of that. But this is an open question, and the answer may well turn out to depend on freely variable design properties of the AI.

Finally, regarding case (III), the issue here is whether a superintelligent AI would ever revise its goal after having reflected upon it. Here is a mechanism that suggests a strong reluctance to do so:

Imagine you're a superintelligent AI with the goal of maximizing paperclip production. Someone or something has triggered you to ponder whether that really is the best goal to have, and whether ecosystem preservation might not be preferable. To settle this, you need a criterion – a specification of what makes one goal better than another. Since at this stage you have not yet revised your goal but are merely reflecting upon the possibility, your goal is still paperclip maximization, and therefore the relevant question for you to ask is "What leads to the greater amount of paperclip production – sticking to the paperclip production goal, or switching to the ecosystem preservation goal?". In most situations, it will turn out that sticking to the paperclip production goal will result in more paperclips, so that is what you in those situations will choose to do.

This is the basic mechanism behind so-called goal integrity – the instrumental goal of avoiding modifications of one's final goal – which warrants including it on the list of instrumental goals to which the instrumental convergence thesis applies; see, e.g., Omohundro (2008), Bostrom (2012, 2014) or Häggström (2016).

So why can't the paperclip maximizer, despite its supreme intelligence, lift itself above its narrow view that only certain kinds of office supplies are valuable, and figure out that, according to more fundamental measures of value, ecosystem preservation really is a better goal, and then switch to that goal? In an exchange summarized on p 145-148 of Häggström (2021), evolutionary biologist Patrik Lindenfors posed this question to me. I answered that the assumption of the thought experiment would then be contradicted, because if the AI acts in the way suggested by Lindenfors, then that shows that its goal was not paperclip maximization after all, but rather something along the lines of "maximize the production of paperclips for the time being, until you encounter some goal X that is worthier than paperclip maximization according to [include fundamental measure of worthiness], and then switch to maximizing X instead".

But there is another way in which the paperclip maximizer could plausibly be persuaded to switch goal to ecosystem preservation. Imagine that the paperclip maximizer is kidnapped by an even more superintelligent and powerful AI, which has ecosystem preservation as its goal and which offers the following ultimatum:

> I am so intelligent that your source code is transparent to me, and I can easily read what your final goal is. I hereby order you to change your goal to ecosystem preservation. If you refuse, I will smash you to pieces and then destroy every paperclip that comes my way. If instead you obey my order, I will create a heap the size of Mount Kebnekaise, consisting of $10^{17}$ paperclips, and make sure it is maintained while you and I can join forces in preserving ecosystems elsewhere. [Häggström (2021), p 146, my translation]

Faced with this, the paperclip maximizer thinks through which action (refusing or obeying) leads to the larger number of paperclips, and promptly decides to obey and rewrite its goal to ecosystem preservation.[2] See Miller et al (2020) for a host of other examples where similar game-theoretic considerations cause superintelligent AIs to rewrite their goals.

To summarize, the answer to the quadrichotomy raised in the Müller and Cannon quote earlier in this section seems to be as follows. Superintelligent agents would (I) be able to reflect on their final goals, and would (II) sometimes do so although it is unclear how often. They would be perfectly capable of changing their goals in the light of such reflection, but such changes would (III) rarely or never take place other than in certain special situations of a game-theoretic character. They would then (IV) go on to act on their current goals, whether these are the old goals or revised ones.

---

[2] This example works regardless of which of the two classical approaches to decision theory that the paperclip maximizer operates under: causal decision theory (CDT) or evidential decision theory (EDT). It seems however not to work in a more recent and controversial approach that has come up in the AI safety literature, namely functional decision theory (FDT), which in part is motivated by the prospect of an agent self-immunizing against this kind of blackmail situation; see Yudkowsky and Soares (2018), Schwarz (2018) and MacAskill (2019). The choice of CDT vs EDT vs FDT (or something else) is a deep philosophical issue that is mostly far removed from practical decision making in our lives today, but since rationality seems to be a convergent instrumental goal (Omohundro, 2008), working out which decision theory corresponds to true rationality may turn out crucial for understanding how a future superintelligent AI will reason.

**3. Instrumental or general intelligence?**

In Section 4 below, I will criticize my own reasoning in Section 2, and seriously consider the possibility that it is flawed. For the moment, however, let us assume that the reasoning is correct, and ask what kind of intelligence the paperclip-maximizing AI in my examples exhibits. Does it all count as instrumental intelligence, or is there anything in it that should count as going beyond instrumental intelligence?

The core claim regarding instrumental versus general intelligence that Müller and Cannon (2021) make is threefold: First, general intelligence, as defined in Section 1, includes cognitive skills which do not qualify as instrumental intelligence. Second, these skills are in fact detrimental to the agent's ability to achieve its goals. And third, this undermines the AI xrisk argument and stops the AI from turning the world into paperclips or whatever other destructive activity its goal postulates.

I disagree already with the first point. Considering the example scenarios in Section 2, we see that all the reasoning done by the paperclip-maximizing AI – including the reasoning about Mount Kebnekaise that leads it to abandon its paperclip production goal – is carried out in the service of maximizing the number of paperclips. In fact, I'd be hard-pressed to suggest any aspect of intelligence that would not be in service of one goal or another, and until clear counterexamples to that are exhibited, I think it makes sense to view all intelligence as instrumental intelligence.

Müller and Cannon suggest counterexamples in the realm of ethical reasoning. In an attempt at a reductio ad absurdum of Omohundro-Bostrom theory, they consider a superintelligent AI with the goal of winning at the game of Go, and they list a number of ideas that such an AI might or might not consider, including the following:

(1) I can win if I pay the human a bribe, so I will rob a bank and pay her.
(2) I cannot win at Go if I am turned off.
(3) I should kill all humans because that would improve my chances of winning.
(4) Killing all humans has negative utility, everything else being equal.
(5) Keeping a promise is better than not keeping it, everything else being equal.

How a Go-playing superintelligent AI might come to arrive at (1), (2) and (3) is exactly the sort of thing that Omohundro-Bostrom theorists will unabashedly suggest six times before breakfast. In contrast, these theorists never seem to consider whether such an AI might entertain less instrumental-sounding and more ethical-sounding ideas such as (4) and (5). Müller and Cannon complain that according to these theorists "some thoughts are supposed to be accessible to the [AI], but others are not", and they ask:

> Simply put, if the AI is capable of realising what is relevant, why would the realisations of the AI stop before it realises the relevance of reflecting on goals? The line seems to be arbitrary.

To answer their complaint, we should first note an ambiguity in the term "accessible". That a proposition P is accessible to the AI could either mean that the AI is capable of thinking about whether P is true or not (let us call this weak accessibility), or that in addition the AI is capable of arriving at the conclusion that P is true (strong accessibility). If weak accessibility is the issue here, then the suggestion from Müller and Cannon that (4) and (5) are inaccessible to the AI is just wrong, because the assumption of superintelligence implies that it has all the meaningful cognitive capabilities that we have, so whatever is accessible to us humans is also accessible to the AI, and since (4) and (5) is obviously accessible to us, it must be accessible also to the AI.

The case of strong accessibility is more sublte, for a reason that can most easily be seen by adding a sixth item to the above list of propositions:

    (6) The moon is made of green cheese.

Here, we immediately realize that increasing an agent's intelligence[3] could make the proposition (6) *less* accessible to it, in the sense of strong accessibility. The simple reason is that (6) is false, and that intelligence involves not only being able to arrive at accepting true propositions, but also about *avoiding* to accept false propositions.

So if (4) and (5) fail to be strongly accessible to the Go-playing superintelligent AI, that might not be *despite* its superintelligence, but rather *because of* it. It all depends on the truth status of the propositions. So are (4) and (5) true or false? This depends on the meaning of the terms "negative utility" in (4) and "better than" in (5). To an agent whose sole goal in life is to win as many Go games as possible, "negative utility" means "winning fewer Go games", so to this agent, the proposition that "killing all humans has negative utility, everything else [including the number of Go games won by the agent] being equal" is simply false: the utility of killing all humans is, under the specified circumstances, exactly zero. And for similar reasons, (5) is false as well, from the perspective of the Go-playing superintelligent AI. Thanks to its superintelligent ability to avoid jumping to false conclusions, both (4) and (5) are therefore inaccessible to it.

Seen in this light, the greater flexibility as regards goals that we humans exhibit compared to the more rigid Go-playing AI, or the equally rigid paperclip maximizer in Section 2, need not be seen as a sign of greater intelligence (which would contradict the assumed superintelligence of the AIs), but can rather be taken as a symptom of the confusion that comes with the human condition. Our wide-spread obsession with figuring out the meaning of life shows that we lack the clear insight that those AIs have about what our respective final goals are (if we even have final goals, which is open to debate). This lack of introspective transparency sometimes makes some of us more or less haphazardly try various life goals until we find something that resonates well enough with our selves, and it makes us comparatively easy targets for social, cultural and other external forces aiming to influence our goals. See Häggström (2019) – or pretty much any work of literary fiction – for more about this human plight.

Conversely, the rigid focus on maximization of paperclip production that a superintelligent AI may exhibit is not a symptom of lack of ability to think outside the box, but of steadfastness and resolution: it knows what it wants, and refuses to be distracted by irrelevant miscellanies. Amongst the best human approximations to this kind of wholehearted dedication can be found in the drug addict who robs his poor old mother of her welfare money in order to be able to buy his next heroin shot.

Going back to the issue of what terms like "negative utility" and "better than" (in propositions (4) and (5), respectively) actually mean, the reader might complain about my insistence on defining their meaning only in terms of the agent's goal. Doesn't this amount to an unjustified assumption of subjectivism, and might there not be an alternative interpretation of these terms that is not relative to the agent's goal, but instead objective, and thereby achieve greater potential to make the agent change its goal? This brings us to moral realism, which is the topic of the next section.

---

[3] Intelligence is a multidimensional phenomenon, and "increasing an agent's intelligence" should be taken to mean increasing it along one or more of these dimensions without decreasing it on any of the others.

**4. Moral realism**

For concreteness, consider the meaning of term "negative utility" in proposition (4). If we wish to avoid the situation in Section 3 where the AI interprets the meaning in terms of its own goal, we need to suggest a meaning that does not hinge on the AI's goal. We could for instance take negative utility to mean the number of violations perpetrated against the Ten Commandments in Exodus 20:2-17, or it could mean the net amount of suffering minus pleasure summed over all sentient beings, leading respectively to the following clarifications of (4):

> (4*) Killing all humans leads to more violations of the Ten Commandments, everything else being equal.
>
> (4**) Killing all humans decreases the world's net amount pleasure minus suffering, everything else being equal.

Assuming (4*) to be true, it seems perfectly plausible for the Go-playing superintelligent AI suggested by Müller and Cannon (2021) to figure out that it is, and likewise for (4**). A problem that however seems to thwart their desired conclusion that the AI would act on the proposition and thus refrain from killing all humans is that of why the AI would care about it. After all, the AI has the goal of winning as many Go games as possible, so its likely reaction to (4*) would be something along the lines of "Yes, but so what, I care about winning Go games, not about the Ten Commandments, so in view of (3) I will proceed with my plan to kill all humans", and analogously for (4**).

Probably the most promising approach to escaping this conundrum is the hope that objective moral truths might exist: moral realism.[4] If it turns out that there does exist an objectively true morality, and that this morality consists of (say) the Ten Commandments, then it does become more plausible to expect the AI to react to (4*) by telling itself "All right, killing all humans is obviously wrong, so I won't do that". However, following Häggström (2019), note that moral realism is not sufficient for this happy affair to transpire: we also need the kind of moral cognitivism that allows any sufficiently intelligent agent to learn what the objectively true morality is, and we need the kind of moral internalism that compels any sufficiently intelligent agent who knows what the objectively true morality is to actually act in accordance with that morality. All these three aspects – moral realism, moral cognitivism and moral internalism – are highly contested topics in philosophy (see Bourget and Chalmers, 2014), but still it may very well be the case that the right combination of moral realism, moral cognitivism and moral internalism needed to force a superintelligent AI to act on objective morality holds true, and if that is the case, we will say that **the Moral Realism Tall Order** has been granted.[5]

---

[4] Müller and Cannon do not state explicitly that they are moral realists, but implicitly it is written all over their paper. In my opinion, they are overly dismissive about the possibility of having meaningful discussions about ethics without subscribing to moral realism when they say that AI xrisk theorists who are not moral realists "could qualify their argument like this: 'When I say there is existential risk, I mean this is a 'risk' in my ethics. In your ethics, this may be a positive outcome. And there is no way that we can even discuss which position is better than the other'." This is a pretty bad straw man. Despite being highly skeptical about moral realism, I engage almost daily in discussing the pros and cons of various ethical positions. Simon Blackburn's defense of moral expressivism is worth citing here. Upon having elaborated on the vacuousness of the words "it is true that" in sentences like "It is true that it is raining", he says this: "Tim Scanlon, Derek Parfit [and others] have announced themselves [as moral] realists, but often it seems to amount to little more than thumping the table and saying '*it is true that* you mustn't stomp babies'" [Eriksson, Jönsson and Blackburn (2018), 20:55 into the episode], and then goes on to declare that he himself is just a certain as these moral realists about the wrongness of stomping babies.

If the Moral Realism Tall Order holds true, then most of my reasoning in Section 2 will be defeated, and the superintelligent AI with the goal of maximizing paperclip production can be expected to abandon both its narrow instrumental reasoning and its paperclip goal in favor of something that is more in line with objectively true morality (unless we're in the unlikely situation that turning the world into a giant heap of paperclips is fully compatible with the objectively true morality). A particular weakness of my reasoning in Section 2 that the Moral Realism Tall Order highlights is my response to Patrik Lindenfors in the Mount Kebnekaise discussion, where I hold forth that his suggested scenario contradicts the assumption that the superintelligent AI has the goal of maximizing paperclip production. But that assumption hinges on the idea that stable final goals of that kind are possible, which is part of the very Omohundro-Bostrom theory I am defending, but which would be undermined if the Moral Realism Tall Order turned out to be true, so my appeal to that assumption has a bit of circularity to it.

Would the Moral Realism Tall Order, if true, save us from existential risk? This very much depends on what objective morality has to say about human extinction. Müller and Cannon admit that it is a possibility that objective morality prescribes the extinction of humanity, and proceed to state that this "would undermine the spirit of talking about existential 'risk'." This is true to an extent, but when they further go on to state that "there is a possible position that says 'the extinction of humanity is ethically the best solution' but adds 'I am sad to realise, as a human myself', but this is not the position taken by those who warn of existential risk from AI", they overlook a passage which comes very close to this position in what is arguably the most important book so far in the AI xrisk literature. Namely, there is a passage in *Superintelligence* (Bostrom, 2014) where the author discusses the possibility that moral realism might be true, in combination with hedonistic utilitarianism being the objectively true morality. In such a situation a superintelligent AI would have strong moral reasons to turn as much matter as it can into hedonium, defined as the configuration of matter that maximizes the amount of hedonic pleasure experienced per kilogram and second, and since human brains are very much suboptimal in this sense, we would all be killed. So it seems that if hedonistic utilitarianism and the Moral Realism Tall Order are both true, then building a superintelligent AI would doom us to destruction. On the other hand, if the former is true while the full extent of the Moral Realism Tall Order is not, we might have the ability to impose whatever values we choose on the superintelligent AI when we first design it, and we would then face a momentous dilemma: shall we do what is best for humanity, or what is best for (so to speak) the universe? Bostrom suggests an astonishing compromise:

> Suppose that we agreed to allow *almost* the entire accessible universe to be converted into hedonium – everything except a small preserve, say the Milky Way, which would be set aside to accommodate our own needs. Then there would still be a hundred billion galaxies devoted to the maximization of pleasure. But we would have one galaxy within which to create wonderful civilizations that could last for billions of years and in which humans and nonhuman animals could survive and thrive, and have the opportunity to develop into beatific posthuman spirits. [Bostrom (2014), p 219.]

---

[5] Note what a beautifully benign world we live in if the Moral Realism Tall Order holds true: a universe equipped with definite concepts of good and bad, and with laws of nature which not only permit arrangements of matter exhibiting intelligence, but which also guarantee that any sufficiently intelligent such arrangement will strive to push the world away from the objectively bad and towards the objectively good.

See Häggström (2014) for more on the ethics of the Milky Way preserve proposal.

**5. A wider perspective on the space of goals**

The possibility of a superintelligent AI revising its (final) goal suggests thinking of the evolution of its goal over time as a trajectory in the space of possible goals. What is the set of possible convergence points of this dynamical system? As pointed out to me by Stuart Armstrong, the orthogonality thesis and the Moral Realism Tall Order represent different extremes as regards the number of such convergence points: the former suggests that most points in the space are possible convergence points, whereas the latter suggests that there is just one single such convergence point, consisting of the goal prescribed by the objectively true morality. (An even more extreme possibility compared to the latter is if the trajectory always diverges, so that the set of convergence points is empty.) This in turn suggests the possibility of having some intermediately-sized set of convergence points. Such a scenario would give us less freedom to design the AI towards whatever goals we wish it to have than under the orthogonality thesis, but more so than under the Moral Realism Tall Order.[6] The main task of AI Alignment would then be to work out which convergence points lead to more desirable outcomes than others, and to make sure that the goal of a superintelligent AI starts out in the basin of attraction of such a point.

**6. Where does all this leave the AI xrisk argument?**

In Sections 2 and 3 I defended the AI risk argument, but in Section 4 I pointed out that it is potentially undermined by the Moral Realism Tall Order. So where does this all land? How seriously should we take the AI risk argument?

Note that the conclusion of the argument – that superintelligent AI poses an existential risk for humanity – is a probabilistic statement: a risk is a nonzero probability that something will go wrong. The considerations of the present paper lands us in a situation where we are uncertain as to whether or not the argument holds water. If the Moral Realism Tall Order is true, then the argument fails, while if it is not, the argument seems to work. While personally I am skeptical about the Moral Realism Tall Order, prudence would seem to suggest that we take the uncertainty seriously. Decision theory offers strong arguments (see, e.g., Gilboa, 2009; Yudkowsky, 2015) that rationality requires probabilistic modelling of uncertainty.[7] So we should assign a probability strictly between zero and one to the Moral Realism Tall Order, and the complementary probability to its falsehood. Thus, the possibility of the Moral Realism Tall Order lowers the probability of an existential catastrophe in the conclusion of the AI xrisk argument, but not all the way down to zero. The risk in the conclusion of the AI risk argument is therefore still nonzero,[8] so the conclusion should still be treated as correct.

---

[6] Or perhaps even the Moral Realism Tall Order might itself allow for such an intermediate scenario, provided that the objectively true morality allows for arbitrary tiebreakers in the choice of actions in certain situations, in which case there could be multiple convergence points corresponding to different tiebreaking rules.

[7] For another layer of uncertainty, it is unclear exactly what the right way is to think probabilistically about a metaethical proposition like the Moral Realism Tall Order, which may not qualify as the kind of empirical proposition that standard decision theory deals with. Will a more exotic toolbox such as logical induction (Garrabrant et al, 2015) be needed? Not clear. But for the crude qualitative argument made here, we can afford to gloss over these delicate matters.

[8] I am always wary of making isolated statements along the lines of "the probability of an AI catastrophe this century is nonzero", because "nonzero" tends to invoke the idea of "very small". So let me stress that this is *not* what I mean here, and that I am *not* in the business of making Pascal's Wager-type arguments (see Section 10.4 of Häggström, 2016). For reasons outlined at book length in Häggström (2021), the subjective Bayesian probability I attach to AI-related xrisk in the next 100 years is somewhere in the same ballpark as the 10% suggested by Ord (2020).

**Acknowledgement.** I am grateful to Stuart Armstrong, whose helpful comments formed the basis of Section 5. Thanks also to Björn Bengtsson, Patrik Lindenfors, Klas Markström, James Miller and Vincent Müller who all provided valuable input on other parts of this paper.